\documentclass[a4paper, amsfonts, amssymb, amsmath, reprint, showkeys, nofootinbib, twoside]{revtex4-1}
\usepackage[english]{babel}
\usepackage[utf8]{inputenc}
\usepackage[colorinlistoftodos, color=green!40, prependcaption]{todonotes}
\usepackage{amsthm}
\usepackage{mathtools}
\usepackage{physics}
\usepackage{xcolor}
\usepackage{graphicx}
\usepackage[left=23mm,right=13mm,top=35mm,columnsep=15pt]{geometry} 
\usepackage{adjustbox}
\usepackage{placeins}
\usepackage[T1]{fontenc}
\usepackage{lipsum}
\usepackage{csquotes}
\usepackage{bm}
\usepackage[pdftex, pdftitle={Article}, pdfauthor={Author}]{hyperref} % For hyperlinks in the PDF
\begin{document}
\title{Fluctuation Relations for Adiabatic Pumping}

\author{Yuki Hino}
    \email[Correspondence email address: ]{yuki.hino@yukawa.kyoto-u.ac.jp}% Your name
\author{Hisao Hayakawa}
    \email[Correspondence email address: ]{hisao@yukawa.kyoto-u.ac.jp}
\affiliation{Yukawa Institute for Theoretical Physics, Kyoto University, Kitashirakawa-oiwake cho, Sakyo-ku, Kyoto 606-8502, Japan}

\date{\today} % Leave empty to omit a date

\begin{abstract}
We derive an extended fluctuation relation for an open system coupled with two reservoirs under adiabatic one-cycle modulation.
We confirm that the geometrical phase caused by the Berry-Sinitsyn-Nemenman curvature in the parameter space generates non-Gaussian fluctuations. This non-Gaussianity is enhanced for the instantaneous fluctuation relation when the bias between the two reservoirs disappears.
\end{abstract}
\maketitle

\section{Introduction} \label{sec:introduction}
Adiabatic pumping is a process where an average current is generated even in the absence of an average bias under slow and periodic modulation of multiple parameters of the system.
The theory of adiabatic pumping was first proposed by Thouless in isolated quantum systems \cite{thouless1, thouless2}. 
He showed that charges can be transported by applying a time-periodic potential to one-dimensional isolated quantum systems under a periodic boundary condition.
He also clarified that the charge transportation in this system is essentially induced by a Berry-phase-like quantity in the parameter space \cite{thouless2, berry, xiao} before Berry proposed the Berry phase \cite {berry}. 
This phenomenon has been observed experimentally in various processes such as charge transport \cite {ex-ch1,ex-ch2,ex-ch2.5,ex-ch3,ex-ch4,ex-ch5,ex-thou1,ex-thou2} and spin pumping \cite {ex-spin1}. 

Later, Brouwer extended the Thouless pumping to open quantum systems \cite{brouwer}.
It was then recognised that the essence of Thouless pumping or geometrical pumping can be described by a classical master equation in which the Berry-Sinitsyn-Nemenman (BSN) phase is the generator of the pumping current \cite{sinitsyn1, sinitsyn2}.
There are various papers on geometrical pumping processes in terms of scattering theory \cite{brouwer, s-th1, s-th2, s-th3,s-th-ch1,s-th-ch2,s-th-ch3,s-th-spin1}, classical master equations \cite {parrondo,usmani,astumian1,sinitsyn1,sinitsyn2,astumian2,rahav,ohkubo,chernyak1,chernyak2} and quantum master equations \cite {qme-spin1,qme-spin2,qme-spin3,qme1,qme2,yuge1}. 
Non-adiabatic pumping processes have also been studied because the pumping current becomes zero in the adiabatic ( i.e., zero frequency) limit \cite {watanabe1, nakajima1}.

We discovered the existence of path-dependent excess entropy production induced by the BSN curvature \cite{sagawa, yuge2, nakajima2}. The existence of the path-dependent entropy in systems under cyclic modulation implies that the direct extension of equilibrium thermodynamics to nonequilibrium processes is not possible, at least for such systems.
Similarly, the geometrical phase effect plays an important role even in heat engines \cite{engine}.
Thus, understanding geometrical pumping is important for both applied and fundamental physics.

In nonequilibrium processes, in addition to the expectation values of physical quantities such as electric and heat currents, their fluctuations are also important. 
Historically, the relationship between fluctuations and responses from base states has been extensively studied, resulting in the Green-Kubo formula for the linear response, the fluctuation-dissipation relation (FDR), Onsager's reciprocity relation etc. \cite{kubo}. 
Furthermore, in the 1990s, the fluctuation theorem (FT) \cite{evans,LL,GC,lebowitz,jar,crooks,esposito2,esposito3,evans-text} was discovered as a relation which holds even in far-from  equilibrium situations. 
The FT expresses the relative probabilities of typical and rare events such as  positive and negative entropy production. 

Let us consider an open system in contact with multiple external reservoirs, which is in a nonequilibrium steady state. In this situation, the probability distribution $P_{\tau}(\hat{J})$ of the current $ \hat{J} $ in time interval $ \tau $  satisfies the steady fluctuation theorem \cite{evans-text,saito1,noh,andrieux}
\begin{equation}\label{SFT}
\lim_{\tau \to \infty} \frac {1} {\tau} \ln \frac {P _{\tau}(\hat{J})} {P_{\tau} (-\hat{J})} = \mathcal{A}^{st}\hat{J},
\end{equation}
where $\mathcal{A}^{st}$ is a steady affinity. For example, when we consider a heat flow, the affinity is given by $\mathcal{A}^{st} = \beta_{R}-\beta_{L} $, where $\beta_{L}$ and $\beta_{R}$ are the inverse temperatures of the left and the right reservoirs, respectively.
Moreover, the FT recovers the Green-Kubo formula, the FDR, Onsager's reciprocity relation and the other nonlinear relations \cite{andrieux}.
The FT in Eq. (1) is a direct consequence of Gaussian fluctuations, since the Gaussian form $P_{\tau}(\hat{J})\sim \exp[-\frac{\tau {\mathcal{A}}^{st}}{4\langle \hat{J}\rangle} (\hat{J}-\langle \hat{J} \rangle)^2]$ with the average current $\langle \hat{J} \rangle$ satisfies Eq. (\ref{SFT}). 

Therefore, systems by non-Gaussian noises do not satisfy the conventional fluctuation theorem \cite{kanazawa,sano}.
Similary, Ren et. al. indicated that the fluctuation theorem is violated in adiabatic pumping because of the existence of the geometrical phase \cite{ren}. 
Watanabe and Hayakawa \cite {watanabe2} analyzed the spin-boson model and verified the violation of the FT in the pumped system. 
Nevertheless, they could not get a concise form of an extended fluctuation theorem for geometric pumping processes. 
We also need to know relations among the cumulants of the current. 
If we can construct the extended fluctuation relation between $P_{\tau}(\hat{J})$ and $P_{\tau}(-\hat{J})$, we can derive the explicit expressions for all cumulants. 

In this paper, we derive two types of fluctuation relations for adiabatic pumping processes by using the generalized master equation with the aid of the full counting statistics (FCS) \cite {esposito2}. 
By using these expressions, we also derive nonequilibrium relations corresponding to the FDR and other key results.
We have confirmed that the geometrical phase generates non-Gaussian fluctuations \cite{kanazawa,sano}, and thus, systems under cyclic modulation do not satisfy the fluctuation theorem. 

The organization of this paper is as follows.
In Sec. \ref{sec:method}, we explain the method used in this paper, the FCS and the generalized master equation. 
In Sec. \ref{sec:pump}, we introduce the adiabatic approximation and show the general form of the cumulants for the pumping current.
Section \ref{sec:FR} is the main part of this paper. 
We calculate the approximate form of the current distribution and derive two types of fluctuation relation.
In Sec. \ref{sec:appli}, we apply our formalism to the spin-boson model to illustrate the role of the non-Gaussian noise in the fluctuation of the pumping current.
Finally, we discuss and summarize our results in Sec. \ref{sec:conclusion}.
In the appendices, we present some detailed calculations to support the description in the main text.

\section{General framework} \label{sec:method}

\subsection{Dynamics}

\begin{figure}[htb]
  \begin{center} 
    \includegraphics[width=8cm]{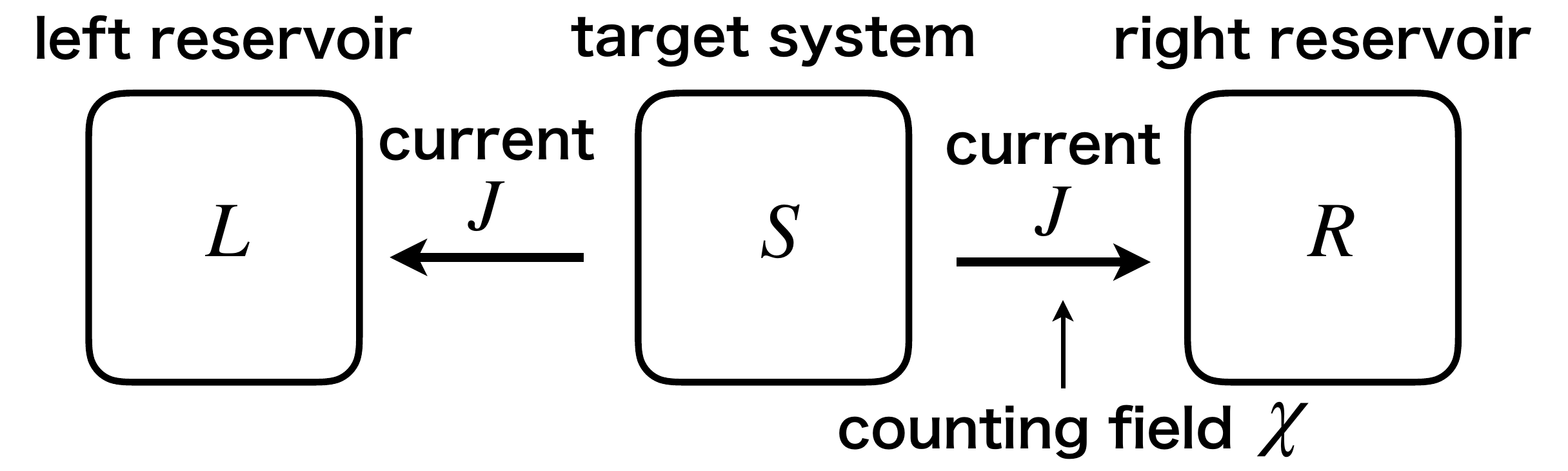}
    \caption{A schematic of the total system which consists of the target system $S$ and the left ($L$) and right ($R$) reservoirs. We measure the current $J$ from $S$ to $R$ by the counting field $\chi$.} 
    \label{system} 
  \end{center}
\end{figure}

In this paper, we consider a total system in which the target system $S$ interacts with two reservoirs $L$ and $R$ (Fig. \ref{system}). 
We assume that the target system $S$ takes discrete $n$ states.  
Let us introduce the vector $|p(t)\rangle := (p_{1}(t), \dots, p_{n}(t))^{T}$, where $p_{i}(t)$ ($1\leq i \leq n$) is the probability that the system takes the state $i$ at time $t$. 
$|p(t)\rangle$ satisfies the normalization condition $\langle 1|p(t)\rangle =1$, where $\langle 1|:=(1,\dots,1)$. 
We assume that the time evolution of $|p(t)\rangle$ is given by the master equation
\begin{align}\label{master}
    \frac{d}{dt}|p(t)\rangle = K(\bm{\alpha}(t))|p(t)\rangle,
\end{align}
where $K(\bm{\alpha}(t))$ is a $n\times n$ matrix characterizing the transition rate of the dynamics with external control parameters $\bm{\alpha}(t)$. 
The $(i,j)$-component of $K(\bm{\alpha}(t))$ is given as $k_{ij}(\bm{\alpha}(t)) := \sum_{\nu=L,R}(k^{\nu}_{ij}(\bm{\alpha}(t)))$, where $k^{\nu}_{ij}(\bm{\alpha}(t))$ ($i \neq j$) is the rate of transition $j \to i$ due to interaction with the reservoir $\nu$ at $t$ and $k^{\nu}_{ii}(\bm{\alpha}(t)) := - \sum_{j,j\neq i} k^{\nu}_{ji}(\bm{\alpha}(t))$.
In this paper, we consider the periodic modulation of the parameters: $\bm{\alpha}(t)=\bm{\alpha}(t+\tau)$ with period $\tau$.

Now, let us introduce the angular frequency $\Omega:=2\pi/\tau$ and the phase $\theta:=\Omega(t-t_{0})$ of parameter modulation, respectively, where $t_{0}$ is a time after which the effect of the initial conditions has become negligible and $|p(t)\rangle$ becomes periodic. 
Then, we rewrite Eq.(\ref{master}) as 
\begin{equation}\label{master-theta}
     \frac{d}{d\theta}|p(\theta)\rangle 
     = \epsilon^{-1} \hat{K}(\bm{\alpha}(\theta))|p(\theta)\rangle,
\end{equation}
where $\epsilon:=\Omega/\Gamma$,  $\hat{K}(\bm{\alpha}(t)):=\Gamma^{-1}K(\bm{\alpha}(t))$, and $\Gamma$ characterizes a typical transition rate between the system and one of the reservoirs. 
Because we are interested in adiabatic modulation, the parameter $\epsilon$ is assumed to satisfiy $\epsilon \ll 1$.

%We next introduce a dimensionless transfer $q_{ij}(t)$ (e.g. the dimensionless heat transfer, the particle transfer, etc.) from the system $S$ to the right reservoir $R$ with the transition $j \to i$ at time $t$. We assume $q_{ij}(t) = -q_{ji}(t)$. 
%The total transfer $q$ in one cycle is determined by the trajectory of the states of the system $S$ 

\subsection{Full counting statistics}

Let us adopt the FCS method \cite{esposito2}. 
FCS enables us to obtain the probability distribution $P(q)$ of the transfer $q$ (e.g. heat transfer, particle transfer, etc.) from the system to a reservoir during one period. 
The scaled cumulant-generating function $G_{\epsilon}(\chi)$ for the transfer $q$ is given by
\begin{equation}\label{cgf}
     G_{\epsilon}(\chi)=\frac{\epsilon}{2\pi} \ln \int_{-\infty}^{\infty}dq P_{\tau}(q)e^{i\chi q},
\end{equation}
where $\chi$ is called the counting field. 
We introduce the counting field only between the system and the right reservoir as shown in Fig. \ref{system}.
To calculate the cumulant-generating function $G_{\epsilon}(\chi)$, we introduce the matrix $\hat{K}(\bm{\alpha}(\theta),\chi)$ which $(i,j)$-component is given as $k_{ij}(\bm{\alpha}(\theta),\chi) := k^{L}_{ij}(\bm{\alpha}(\theta)) + k^{R}_{ij}(\bm{\alpha}(\theta)) e^{i\chi q_{ij}(\theta)}$, where $q_{ij}(\theta)$ is the transfer from the system $S$ to the right reservoir $R$ with the transition $j \to i$ at $\theta$ \cite{sinitsyn1,sinitsyn2}. 
We assume $q_{ij}(\theta) = -q_{ji}(\theta)$. 
Let us consider the time evolution of vector $|p(\theta,\chi)\rangle$ discribed by the generalized master equation
\begin{equation}\label{gcme}
     \partial_{\theta}|p(\theta,\chi)\rangle 
     = \epsilon^{-1} \hat{K}(\bm{\alpha}(\theta),\chi)|p(\theta,\chi)\rangle,
\end{equation}
with initial condition $|p(0,\chi)\rangle = |p(0)\rangle$. 
By using the solution of Eq. (\ref{gcme}), the scaled cumulant-generating function can be written as
\begin{align}
    G_{\epsilon}(\chi) = \frac{\epsilon}{2\pi} \ln\langle 1|p(\theta,\chi)\rangle.
\end{align}
The $n$-th cumulant for the transfer $q$ can be calculated by the $n$-th derivative of $G_{\epsilon}(\chi)$ as
\begin{equation}\label{n-th-cumulant}
    \frac{\epsilon}{2\pi}\langle q^{n} \rangle_{c} = 
    \left.\frac{\partial^{n}}{\partial(i\chi)^{n}}G_{\epsilon}(\chi) \right|_{i\chi=0}.
\end{equation}
Note that the average of the current $J:=\Omega q/(2\pi\Gamma)=\epsilon q/2\pi$ can be written as
\begin{align}
    \langle J \rangle = \frac{\epsilon}{2\pi} \langle q \rangle_{c} = \left.\partial_{i\chi} G_{\epsilon}(\chi)\right|_{\chi=0}.
\end{align}

\section{Adiabatic pumping} \label{sec:pump}

In this section, we briefly review the method to obtain the cumulant-generating function for the transfer $q$ under the adiabatic process \cite{ren,watanabe2,affinity}.
First, we adopt the adiabatic approximation
\begin{align}\label{ad}
     |p(\theta,\chi)\rangle 
     \simeq
     &\exp\left\{\frac{1}{\epsilon}\int_{0}^{\theta} d\theta' [\lambda(\theta',\chi) - \epsilon\; v(\theta',\chi)]\right\}
\notag \\
& \times |r(\theta,\chi) \rangle
\langle 1 |r(0,\chi)\rangle^{-1},
\end{align}
where $\lambda(\theta,\chi)$ is the eigenvalue of $\hat{K}(\bm{\alpha}(\theta),\chi)$, which reduces to zero in the limit $\chi\to 0$.
Here we define $v(\theta,\chi):=\langle l(\theta,\chi)|\partial_{\theta}|r(\theta,\chi)\rangle$, where $\langle l(\theta,\chi)|$ and $|r(\theta,\chi)\rangle$ are the left and right eigenvectors corresponding to $\lambda(\theta,\chi)$, respectively. 

The scaled cumulant-generating function for the transfer $q$ is given by
\begin{equation}
    G_{\epsilon}(\chi)=\frac{\epsilon}{2\pi} \ln\langle 1|p(2\pi,\chi)\rangle.
\end{equation}
By using the adiabatic solution (\ref{ad}), we obtain
\begin{equation}\label{cgf-ad}
G_{\epsilon}(\chi) =\Lambda(\chi) - \epsilon V(\chi),
\end{equation}
where
\begin{equation}
    \Lambda(\chi):=\frac{1}{2\pi}\int_{0}^{2\pi}d\theta \lambda(\theta,\chi)
\end{equation}
is the dynamical part and 
\begin{align}
    V(\chi):=&\frac{1}{2\pi}\int_{0}^{2\pi}d\theta v(\theta,\chi) \notag \\
    =& \frac{1}{2\pi}\iint_{S} d\alpha_{m}d\alpha_{n} \mathcal{F}(\bm{\alpha},\chi),
\end{align}
is the geometrical part.
Here we define $\mathcal{F}_{mn}(\bm{\alpha},\chi):= \partial_{\alpha_{m}}\langle l_{0}(\bm{\alpha},\chi)| \wedge \partial_{\alpha_{n}} |r_{0}(\bm{\alpha},\chi) \rangle$ and $S$ is the open surface enclosed by the contour $C$ of parameter control (see Fig. \ref{parameter}). 
The derivation of Eqs. (\ref{ad}) is given in Appendix \ref{sec:ad-app}.
Note that $\lambda(\theta,\chi)$ satisfies the Levitov-Lesovik-Gallavotti-Cohen (LLGC) symmetry $\lambda(\theta,\chi)=\lambda(\theta, -\chi + i\mathcal{A}(\theta))$ with the instantaneous bias $\mathcal{A}(\theta)$ (e.g. inverse temperature difference $\beta_{R}(\theta)-\beta_{R}(\theta)$ in the case of heat current) while $v(\theta,\chi)$ does not satisfy such a symmetry, i.e. $v(\theta,\chi)\neq v(\theta, -\chi + i\mathcal{A}(\theta))$ \cite{ren,watanabe2,affinity}.

\begin{figure}[htb]
  \begin{center} %センタリングする
    \includegraphics[width = 3cm]{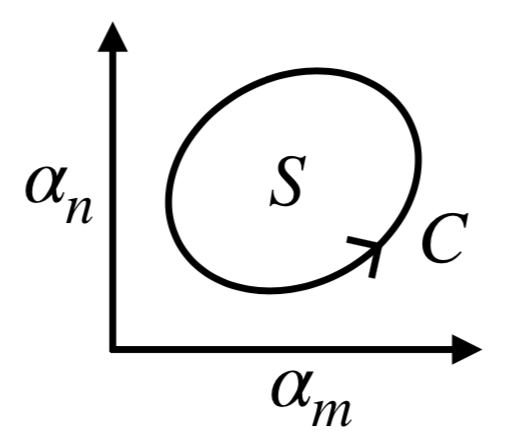}
    \caption{A schematic of a contour $C$ and the surface enclosed by $C$ in the parameter space spanned by $(\alpha_{m},\alpha_{n})$.} %タイトルをつける
    \label{parameter} %ラベルをつけ図の参照を可能にする
  \end{center}
\end{figure}

\section{Fluctuation relations} \label{sec:FR}
This section is the main part of this paper. In this section, we present the general expressions for two types of fluctuation relations for adiabatic pumping processes. In subsection \ref{subsec:cycle}, we discuss the cyclic fluctuation relation and in subsection \ref{subsec:inst}, we give the general expression for the instantaneous fluctuation relation.
   
\subsection{Cyclic fluctuation relation}
\label{subsec:cycle}
Because of Eqs. (\ref{cgf}) and (\ref{cgf-ad}) the probability distribution function $P_{\epsilon}(J)$ of the current $J=\epsilon q/2\pi$ under one-cycle modulation with the parameter $\epsilon$ is given by
\begin{align}
    P_{\epsilon}(J) &= \frac{2\pi}{\epsilon} P_{\tau}(q) \notag \\
    &= \frac{2\pi}{\epsilon}\int_{-\infty}^{\infty}\frac{d\chi}{2\pi}\; e^{-\frac{2\pi}{\epsilon}[i\chi J - G_{\epsilon}(\chi)]}
    \notag \\
    &= \frac{1}{\epsilon}\int_{-\infty}^{\infty} d\chi\;
    e^{-\frac{2\pi}{\epsilon}[i\chi J - \Lambda(\chi)]-V(\chi)} .
\end{align}
When we consider an adiabatic pumping process ($\epsilon \ll 1$), the contribution of  $V(\chi)$ is small. By using the saddle point approximation, $P_{\epsilon}(J)$ can be evaluated as
\begin{equation}\label{pdf}
    P_{\epsilon}(J) 
    \simeq \frac{1}{\sqrt{\epsilon\Lambda^{(2)}(\chi_{c}(J))}}
    e^{-\frac{2\pi}{\epsilon}[I(J)+\epsilon V(\chi_{c}(J))]},
\end{equation}
where we have introduced the large deviation function (LDF)
\begin{equation}\label{ldf}
    I(J):= i\chi_{c}(J) J - \Lambda(\chi_{c}(J)),
\end{equation}
where $\chi_{c}(J)$ is the saddle point which satisfies $\left. \partial_{i\chi}\Lambda(\chi) \right|_{\chi=\chi_{c}(J)} =J$ and $\Lambda^{(2)}(\chi_{c}(J)):= \left. \partial_{i\chi}^{2}\Lambda(\chi)\right|_{\chi=\chi_{c}(J)}$. 
It is expected that $I(J)$ satisfies the symmetry relation
\begin{equation}\label{sym-I}
    I(J)-I(-J) = -\mathcal{A}J
\end{equation}
where $\mathcal{A}$ is the dynamical affinity, which is determined by quantities of the left and the right reservoir. In fact, it was confirmed numerically that $\mathcal{A}$ is given as
\begin{equation}\label{fermion-affinity}
\mathcal{A} = \ln \frac{\int_{0}^{2\pi} d\theta n^{\pm}_{R}(\theta)  (1 \pm n^{\pm}_{L}(\theta)) }{\int_{0}^{2\pi} d\theta n^{\pm}_{L}(\theta)  (1 \pm n^{\pm}_{R}(\theta)) },
\end{equation}
in two-level systems of fermions \cite{affinity} and bosons (see Appendix \ref{spin-boson-sym}).
Here $n^{\pm}_{\nu}(\theta)$ is the Bose ($+$) and Fermi ($-$) distribution of the $\nu$-th reservoir, respectively.
Because of the absence of the LLGC symmetry for $v(\theta,\chi)$, it is obvious that $V(\chi)$ does not have the corresponding symmetry.
By using Eqs. (\ref{pdf}) and (\ref{sym-I}), we obtain the cyclic fluctuation relation
\begin{align}\label{FR1}
\frac{\epsilon}{2\pi}\ln\frac{P_{\epsilon}(J)}{P_{\epsilon}(-J)}
= &\mathcal{A}J - \epsilon[ V(\chi_{c}(J))-V(\chi_{c}(-J))] 
\notag \\
&- \frac{\epsilon}{4\pi}\ln\frac{\Lambda^{(2)}(\chi_{c}(J))}{\Lambda^{(2)}(\chi_{c}(-J))}.
\end{align}
This is one of our main results. 
The second term on the right hand side of Eq. (\ref{FR1}) stands for the geometrical phase contribution which is much smaller than the first term. 
When $V(\chi_{c}(J))=0$, Eq. (\ref{FR1}) reduces to the steady fluctuation theorem in driven systems \cite{affinity}. 
Thus, Eq. (\ref{FR1}) can be regarded as an extension of the fluctuation theorem for the adiabatic pumping process. 
If the trajectory $C$ of the parameter modulation is symmetric with respect to the parameters $\alpha_{m}$ and $\alpha_{n}$,  the average bias is zero, i. e. $\mathcal{A}=0$ and $V(-\chi)=-V(\chi)$, $\Lambda(\chi)=\Lambda(-\chi)$. In this case, Eq. (\ref{FR1}) is reduced to
\begin{equation}\label{FR2}
    \ln\frac{P_{\epsilon}(J)}{P_{\epsilon}(-J)}
    =  - 4\pi V(\chi_{c}(J)),
\end{equation}
which can be expressed only by the geometrical phase. As will be shown in Sec. \ref{sec:appli}, Eq. (\ref{FR2}) contains contributions nonlinear in $J$.

\subsection{Instantaneous fluctuation relation}
\label{subsec:inst}
In this subsection, let us consider the instantaneous fluctuation relation of our system. 
If the master equation (\ref{gcme}) does not contain any singularities, the cumulant generating function can be written as
\begin{equation}\label{cgf-discrete}
    G_{\epsilon}(\chi)=\frac{1}{2\pi}\int_{0}^{2\pi} d\theta g(\theta,\chi)
    =\lim_{N\to\infty} \frac{1}{N}\sum_{n=1}^{N} g_{n}(\chi),
\end{equation}
where $g(\theta,\chi) := \lambda(\theta,\chi) - \epsilon v(\theta,\chi)$ is the instantaneous cumulant-generating function. Here we discretize $\theta$  in the interval $[0,2\pi]$ as in the last expression of Eq. (\ref{cgf-discrete}), where we have introduced $\theta_{n}:= n \Delta\theta$, $\Delta\theta := 2\pi/N$, 
$g_{n}(\chi):=g(\theta_{n},\chi)$, $\lambda_{n}(\chi):=\lambda(\theta_{n},\chi)$, $v_{n}(\chi):=v(\theta_{n},\chi)$.
From Eq. (\ref{cgf-discrete}), the distribution of the current during one cycle can be decomposed into
\begin{align}\label{decomposition}
    &P_{\epsilon}(J) \notag \\
    &=  
    \lim_{N\to\infty} N \int_{-\infty}^{\infty}\prod_{n=1}^{N} dJ_{n} \delta\left(J-\frac{1}{N}\sum_{n=1}^{N} J_{n} \right) \prod_{n=1}^{N} p_{n}(J_{n}),
\end{align}
where
\begin{equation}\label{pdf-n}
    p_{n}(J_{n}) := \int_{-\infty}^{\infty} \frac{d\chi}{\epsilon N} e^{-\frac{2\pi}{\epsilon N}[i\chi J_{n}- g_{n}(\chi)]}
\end{equation}
is the instantaneous distribution at of the current $J_{n}$ at $\theta =\theta_{n}$.
(The derivation of Eqs. (\ref{decomposition}) and (\ref{pdf-n}) is explained in Appendix \ref{sec:decomposition}).
Here we assume $(\epsilon N)^{-1}\gg 1$. This means that $(\epsilon N)^{-1}$ is enough large to relax the system to the instantaneous steady state.

By an argument parallel to that used in the previous subsection, the instantaneous distribution for the current $J_{n}$ is given as
\begin{equation}\label{ins-pdf}
p_{n}(J_{n})
\simeq
\frac{1}{\sqrt{\epsilon N \lambda_{n}^{(2)}(\chi_{c}(J_{n}))}}
e^{ -\frac{2\pi}{\epsilon N}[I_{n}(J_{n}) +v_{n}(\chi_{c}(J_{n}))]},
\end{equation}
where  $\lambda_{n}^{(2)}(\chi_{c}(J_{n})):=\left.\partial_{i\chi}^2\lambda_{n}(\chi)\right|_{\chi=\chi_{c}(J)}$ and
we have introduced the instantaneous LDF
\begin{equation}\label{ldf-n}
    I_{n}(J_{n}):= i\chi_{c}(J_{n}) J_{n}-\lambda_{n}(\chi_{c}(J)),
\end{equation}
where $\chi_{c}(J_{n})$ satisfies $\left. \partial_{i\chi} \lambda_{n}(\chi)\right|_{\chi=\chi_{c}(J_{n})} =J_{n}$. 
Because the instantaneous eigenvalue $\lambda_{n}(\chi)$ satisfies the LLGC symmetry \cite{LL,GC,lebowitz} $\lambda_{n}(\chi)=\lambda_{n}(-\chi+i \mathcal{A}_{n})$, $I_{n}(J)$ satisfies the symmetry relation
\begin{equation}\label{ins-sym}
    I_{n}(J_{n})-I_{n}(-J_{n})=-\mathcal{A}_{n}J_{n},
\end{equation}
where $\mathcal{A}_{n}$ is the instantaneous affinity, which is given by, for example, $\mathcal{A}_{n}=\beta_{R}(\theta_{n})-\beta_{L}(\theta_{n})$ when we control the inverse temperatures $\beta_{L}$ and $\beta_{R}$ of the left and the right reservoirs. 

From Eqs. (\ref{ins-pdf}) and (\ref{ins-sym}), we obtain the instantaneous fluctuation relation
\begin{align}\label{ins-FR}
    \lim_{\epsilon N\to 0}
    &\frac{\epsilon N}{2\pi} \ln\frac{p_{n}(J_{n})}{p_{n}(-J_{n})} = \mathcal{A}_{n}J_{n} \notag \\
    &- \epsilon [v_{n}(\chi_{c}(J_{n}))-v_{n}(\chi_{c}(-J_{n}))] \notag \\
    &-\frac{\epsilon N}{4\pi}\ln\frac{\lambda^{(2)}(\chi_{c}(J))}{\lambda^{(2)}(\chi_{c}(-J))}.
\end{align}
The second term on the right hand side of Eq. (\ref{ins-FR}) expresses the geometrical phase effect at $\theta=\theta_{n}$, which is much smaller than the first term. If $\mathcal{A}_{n}=0$, the first and third terms vanish then the geometrical contribution becomes dominant. As will be shown in Sec. \ref{sec:appli}, the geometric contribution of Eq. (\ref{ins-FR}) is a nonlinear function of $J_{n}$.

\section{Application to the spin-boson system} \label{sec:appli}
The results presented in the previous section can be used for an arbitrary adiabatic pumping process if the process can be described by the master equation Eq. (\ref{gcme}). To know the explicit contribution of the geometric phase in the extended fluctuation relations such as  Eqs. (\ref{FR1}), (\ref{FR2}) and (\ref{ins-FR}), we need to know the details of the eigenstates and the eigenvalues of the operator $\hat{K}(\bm{\alpha}(\theta),\chi)$. 
Here, we apply the general results of Sec. \ref{sec:FR} to the spin-boson model \cite{breuer}. 

\subsection{The spin-boson model}

\begin{figure}[htb]
  \begin{center} 
    \includegraphics[width=8cm]{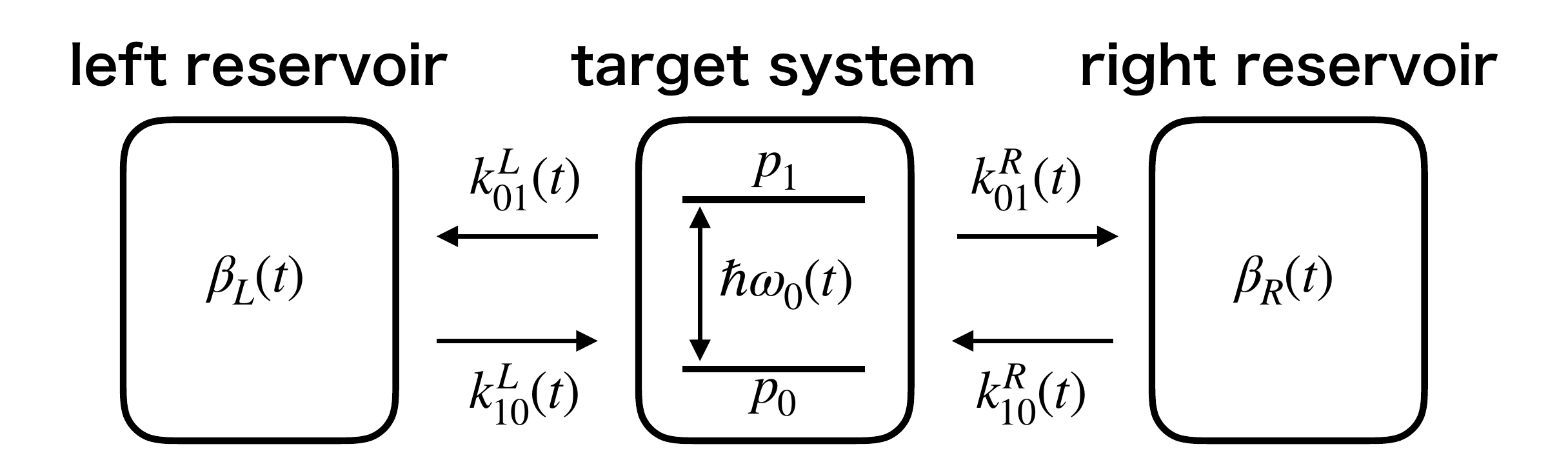}
    \caption{A schematic of the spin-boson model.} 
    \label{spin-boson} 
  \end{center}
\end{figure}

In this section, we consider a single spin coupled to two bosonic reservoirs with inverse temperature $\beta_{\nu}$ ($\nu=L,R$). 
We note the vector $|p(t)\rangle = (p_{0}(t),p_{1}(t))$, where $p_{0}$ and $p_{1}$ are the probability to take the dowm and up state, respectively.
In this model, the transition matrix $K(t)$ in Eq. (\ref{master}) is a $2\times 2$-matrix and its component is given as
\begin{align}\label{k01-t}
    &k^{\nu}_{01}(\bm{\alpha}(t)) := \Gamma_{\nu}(t) (n_{\nu}(\omega_{0}(t),\beta_{\nu}(t))+1), \\
    \label{k10-t}
    &k^{\nu}_{10}(\bm{\alpha}(t)) := \Gamma_{\nu}(t) n_{\nu}(\omega_{0}(t),\beta_{\nu}(t)),
\end{align}
where $n_{\nu}(\omega_{0}(t),\beta_{\nu}(t)):=(e^{\beta(t)\hbar\omega_{0}(t)}-1)^{-1}$ the Bose distribution function in the $\nu$-th reservoir, $\hbar\omega_{0}$ is the energy difference between up and down states.
$\Gamma_{\nu}$ is the transition rate between the system and the $\nu$-th reservoir.

Now, the dimensionless forms of Eqs. (\ref{k01-t}) and (\ref{k10-t}) are written as
\begin{align}
    &\hat{k}^{\nu}_{01}(\bm{\alpha}(\theta)) := \gamma_{\nu}(\theta) (n_{\nu}(\omega_{0}(\theta),\beta_{\nu}(\theta))+1), \\
    &\hat{k}^{\nu}_{10}(\bm{\alpha}(\theta)) := \gamma_{\nu}(\theta) n_{\nu}(\omega_{0}(\theta),\beta_{\nu}(\theta)),
\end{align}
where $\gamma_{\nu}(\theta):=\Gamma_{\nu}(\theta)/\Gamma$ with  $\Gamma:=\sum_{\nu}\int^{2\pi}_{0}d\theta \Gamma_{\nu}(\theta)$. 
Note that $\epsilon:=\Omega/\Gamma$ is held in this case. 
The transfer corresponding the transition $1\to 0$ is given as $q_{01}=\hbar \omega_{0} =-q_{10}$.
We consider the set of parameter $\bm{\alpha}=\{\omega_{0},\gamma_{\nu},\beta_{\nu}\}_{\nu=L,R}$.

\subsection{Cyclic fluctuation relation for the spin-boson model}

\begin{figure}[htb]
  \begin{center} %センタリングする
   \includegraphics[width=8cm]{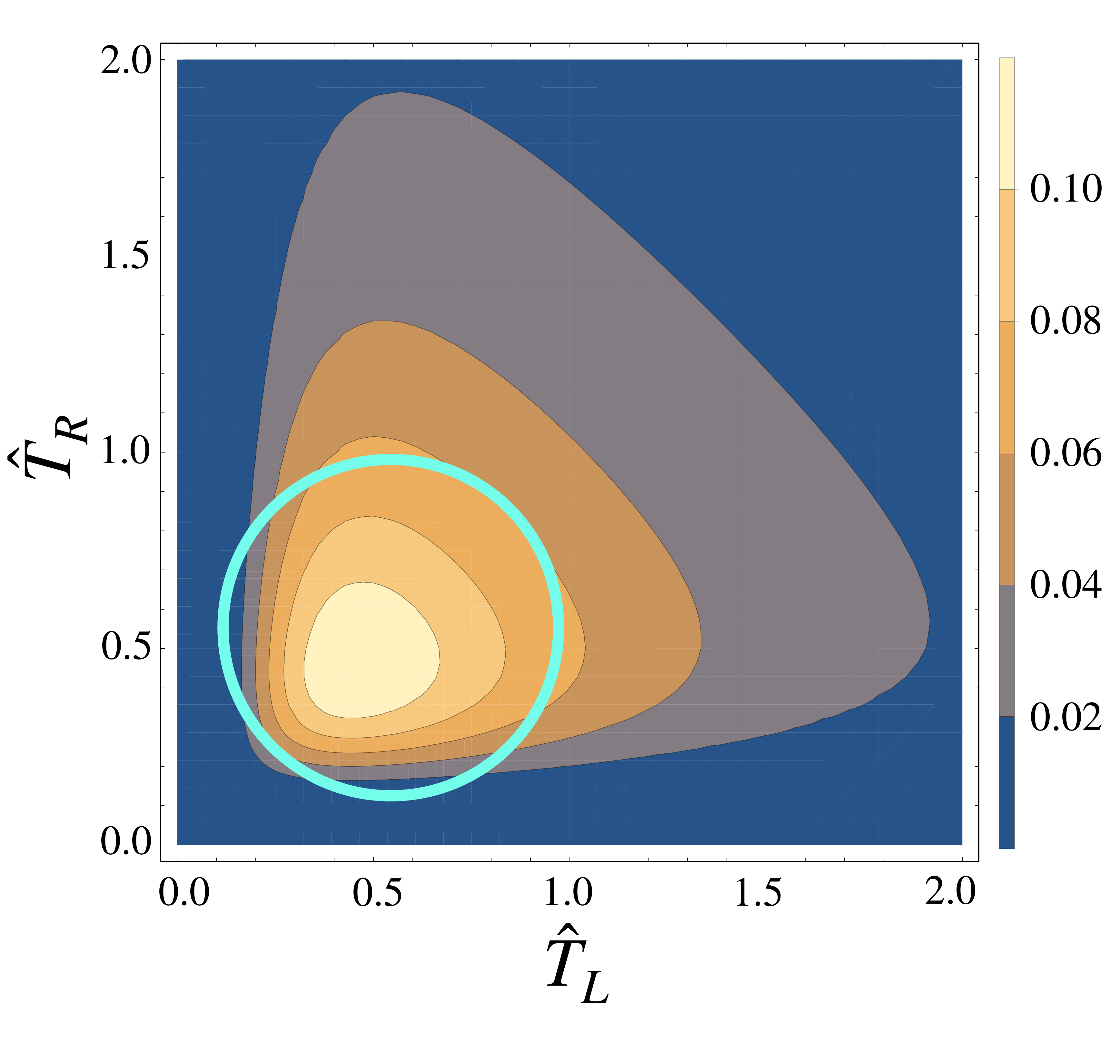}
   \includegraphics[width=8cm]{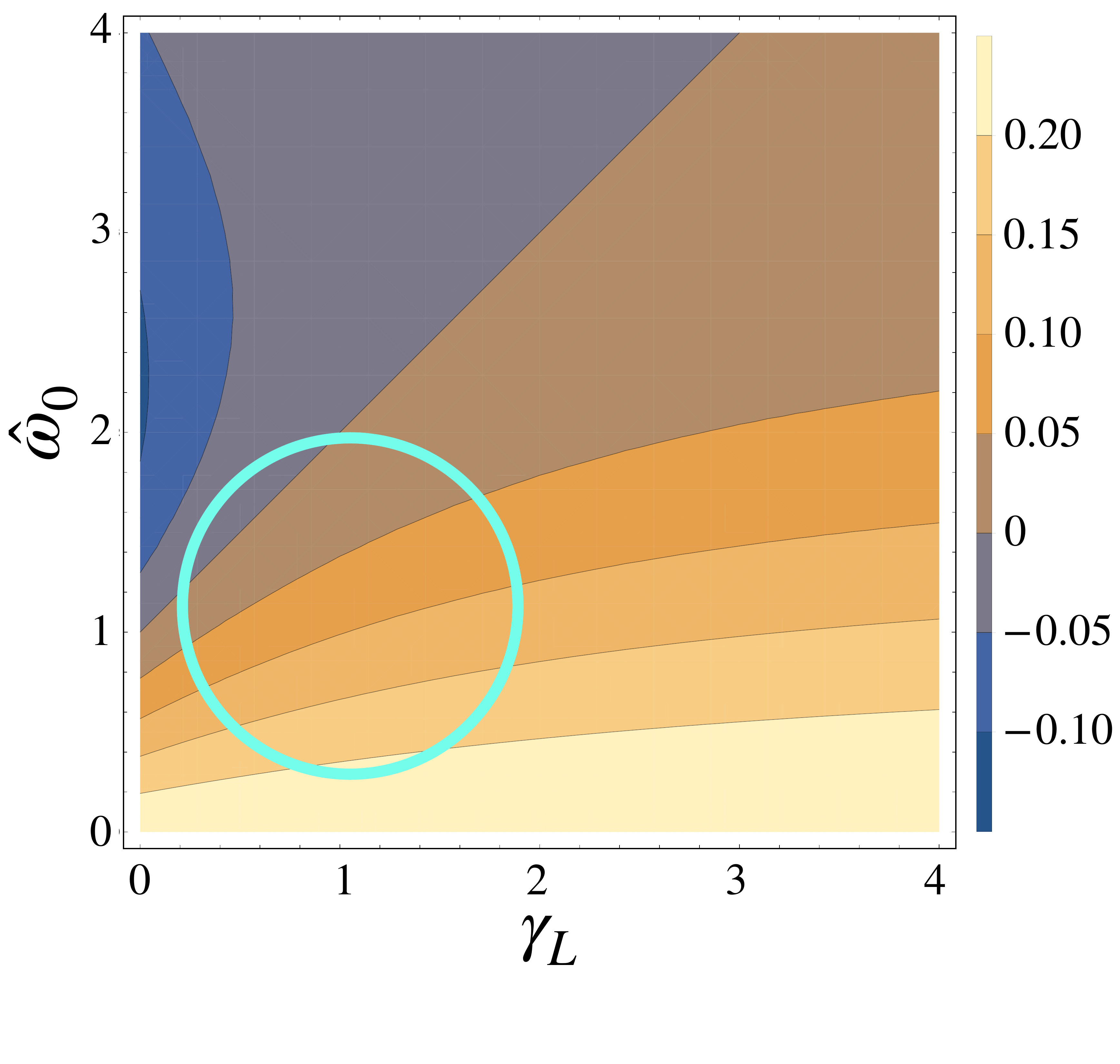}
   \caption{Plots of the BSN curvatures $F_{mn}(\bm{\alpha})$ in Eq.(\ref{BSNcurvature}) in the parameter space in the first case (upper) and the second case (lower). The cyan circle represents the trajectory of the parameter modulation. The geometrical current is determined by the integral of $F_{mn}(\bm{\alpha})$ in the area surrounded by this circle.} %タイトルをつける
   \label{BSN} %ラベルをつけ図の参照を可能にする
  \end{center}
 \end{figure}

Let us calculate the right hand side of Eq. (\ref{FR1}) for two types of modulations. 
In the first case, we control the temperature of the left and right reservoirs as
\begin{align}
\label{betaL}
&\hat{T}_{L}(\theta):=(\hbar\omega_{0}\beta_{L}(\theta))^{-1}=\hat{T}_{0}+\hat{T}_{A}\cos (\theta +\pi/4), \\
\label{betaR}
&\hat{T}_{R}(\theta):=(\hbar\omega_{0}\beta_{R}(\theta))^{-1}=\hat{T}_{0}+\hat{T}_{A}\sin (\theta +\pi/4).
\end{align}
where $\hat{T}_{0}$ and $\hat{T}_{A}$ are the center and the amplitude of the dimensionless temperatures $\hat{T}_L$ and $\hat{T}_R$, respectively\footnote{Of course, the continuous control of the temperatures is not easy but possible as follows.
(i) For example, an effective temperature is continuously changed in Ref. \cite{izumida}. 
(ii) Since we consider an adiabatic process, it is possible to replace a  reservoir by another reservoir having a different temperature and wait for the system to relax to a steady state. If we can repeat this process, we can change the temperatures at the reservoirs as in Eqs. (\ref{betaL}) and (\ref{betaR}).}.
For simplicity, we assume $\Gamma_{L}=\Gamma_{R}=\Gamma$.
In the second case, we control the dimensionless line-width between the target system and the left reservoir and the energy level in the target system such that
\begin{align}
&\gamma_{L}(\theta)=\gamma_{R}+\gamma_{A}\cos(\theta), \\
&\hat{\omega}_{0}(\theta):=\beta \hbar\omega_{0}(\theta)=\hat{\omega}_{C} +\hat{\omega}_{A}\sin(\theta).
\end{align}
where $\gamma_{A}$ is the amplitude of the dimensionless line-width $\gamma_{L}$. 
$\hat{\omega}_{C}$ and $\hat{\omega}_{A}$ are the center and the amplitude of the dimensionless energy gap between two levels in the target system
\footnote{It is easy to control $\gamma_L(\theta)$ and $\omega_0(\theta)$  in experiments \cite{ex-ch1,ex-ch2,ex-ch2.5,ex-ch3,ex-ch4,ex-ch5}.}. 
For simplicity, we assume $\beta_{L}=\beta_{R}=\beta$. 
The BSN curvature $F_{mn}(\bm{\alpha})$ introduced in Eq. (\ref{BSNcurvature}) of the first case in Sec.\ref{sec:appli} is given as
\begin{equation}
F(\hat{T}_{L},\hat{T}_{R})=\frac{1}{8}\frac{n_{L}^{2}n_{R}^{2}}{\hat{T}_{L}^{2}\hat{T}_{R}^{2}(1+n_{L}+n_{R})^{3}},
\end{equation}
where $n_{\nu}$ ($\nu=L$ or $R$) is $n_{\nu}:= (e^{\beta_{\nu}\hbar\omega_{0}}-1)^{-1}$.
Similarly, the BSN curvature of the second case in Sec.\ref{sec:appli} is given as 
\begin{equation}
F(\gamma_{L},\hat{\omega}_{0})=\frac{(\gamma_{L}+1-\hat{\omega}_{0})n_{L}(1+n_{L})}{(\gamma_{L}+1)(1+2n_{L})^{2}}.
\end{equation}
Their plots are given in Fig.\ref{BSN}.
In both cases, because the affinity satisfies $\mathcal{A}=0$, the geometrical phase effect $V(\chi_{c}(J))$ plays an important role as in Eq.  (\ref{FR2}).

\begin{figure}[htb]
  \begin{center} %センタリングする
    \includegraphics[width=8cm]{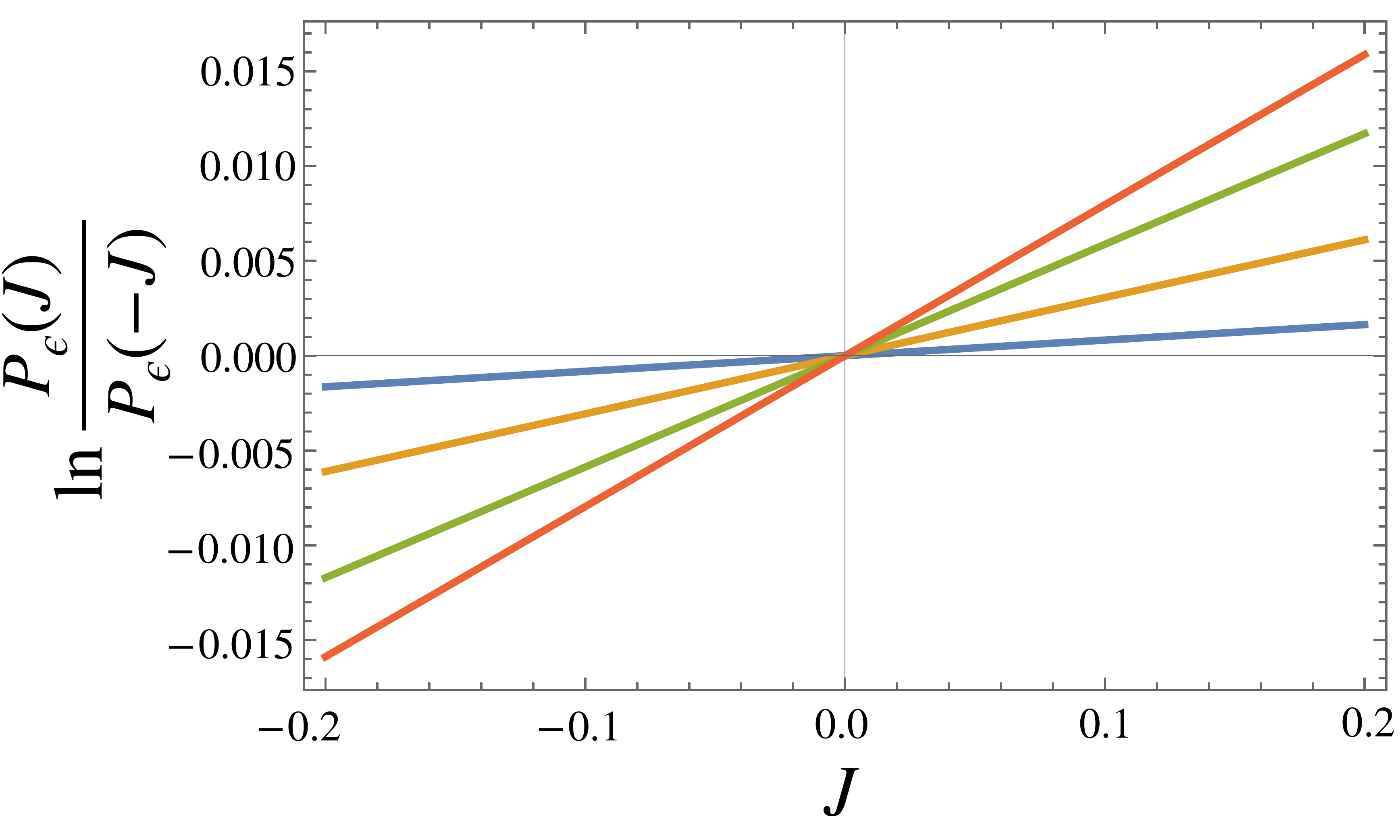}
    \includegraphics[width=8cm]{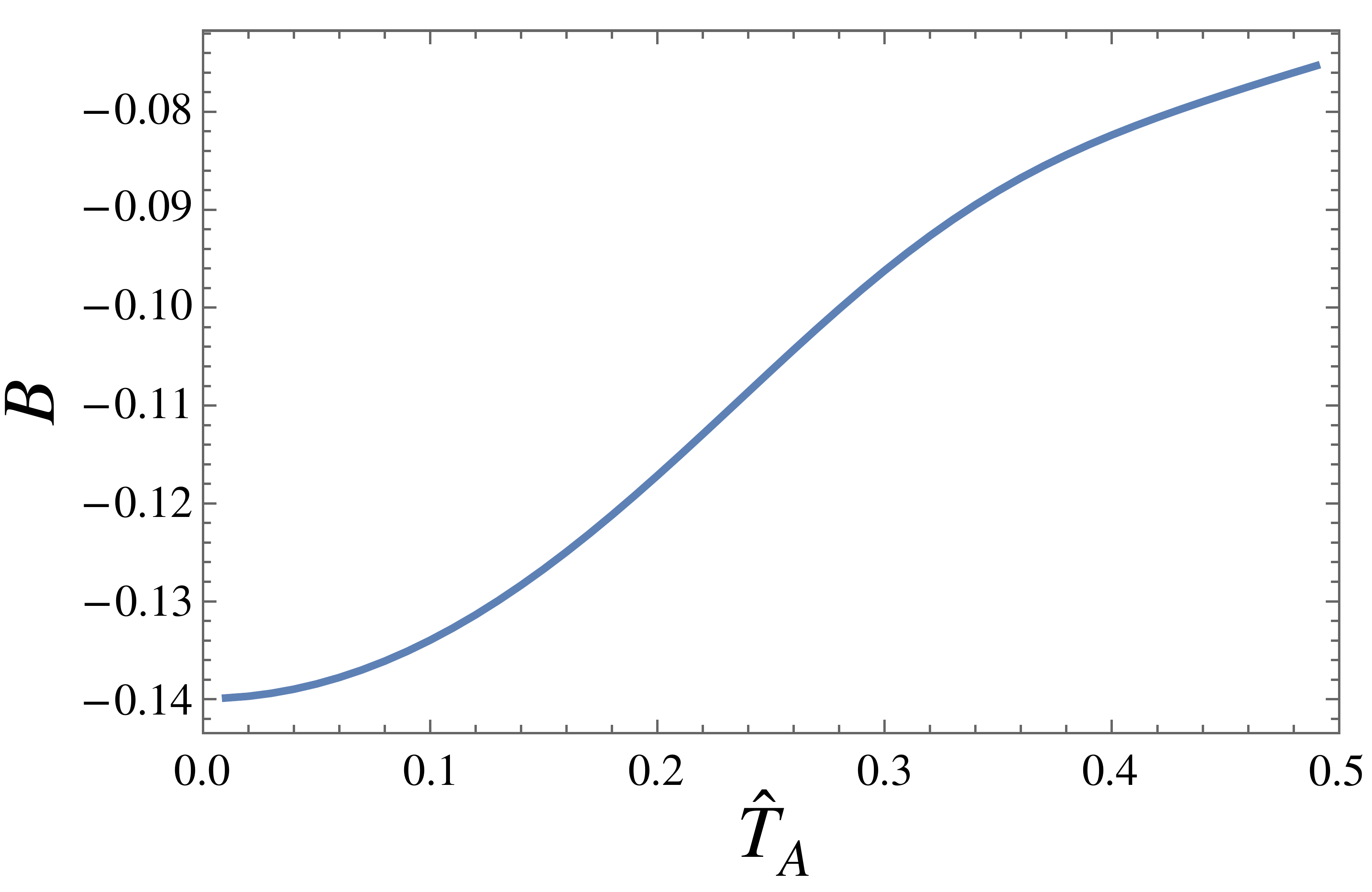}
 \caption{Cyclic fluctuation relation under the control of the reservoir temperatures ( the first case in the main text ). In the top figure, we plot the lines at $\hat{T}_{0}=0.5$, $\hat{T}_{A}=0.1,0.2,0.3,0.4$ (colors are blue, yellow, green and red, respectively) and $\epsilon = 0.001$ as an example of the first case. In the bottom figure, we plot $\hat{T}_{A}$ dependence of the cubic contribution $B_{C}$  at $\hat{T}_{0}=0.5$.} %タイトルをつける
    \label{FT-beta} %ラベルをつけ図の参照を可能にする
 \end{center}
\end{figure} 
 
\begin{figure}[htb] 
  \begin{center} %センタリングする
    \includegraphics[width=8cm]{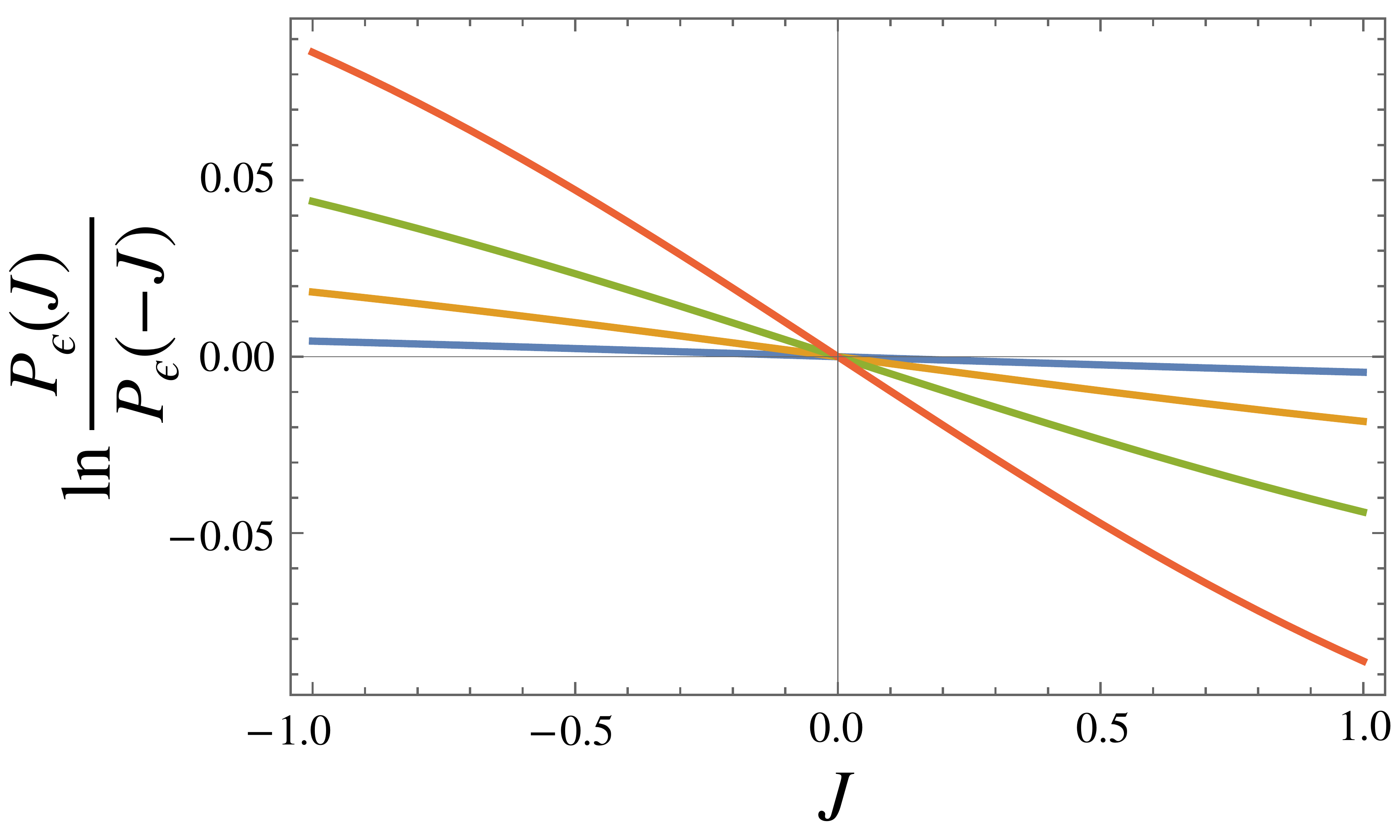}
    \includegraphics[width=8cm]{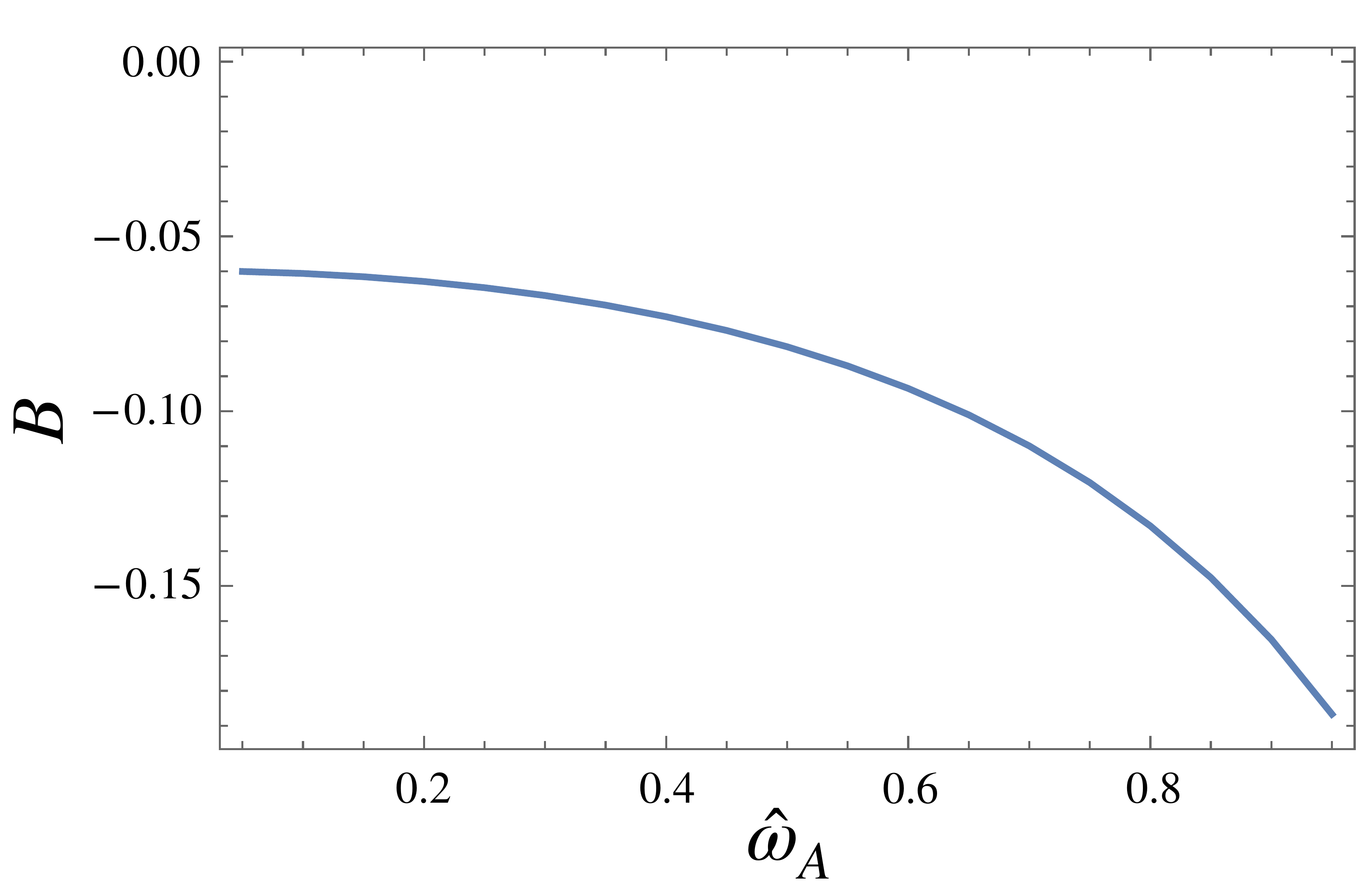}
  \caption{Cyclic fluctuation relation under the control of the energy level and left line-width for $\hat{\omega}_{C}=1$, $\hat{\omega}_{A}= 0.2,0.4,0.6,0.8$ (blue, yellow, green and red, respectively), $\gamma_{R} = 1$, $\gamma_{A}=0.2,0.4,0.6,0.8$ (blue, yellow, green and red, respectively) and $\epsilon = 0.001$ as an example of the second case. In the bottom figure, we plot $\hat{\omega}_{A}$($=\gamma_{A}$) dependence of the cubic contribution $B_{C}$  at $\hat{\omega}_{C}=\gamma_{R}=1$.} %タイトルをつける
    \label{FT-gw} %ラベルをつけ図の参照を可能にする
 \end{center}
\end{figure}

By expanding the right hand side of Eq. (\ref{FR2}) with respect to $J$, we obtain
\begin{equation}\label{FR-SB}
\ln\frac{P_{\epsilon}(J)}{P_{\epsilon}(-J)} \simeq A_{C}[J + B_{C} J^{3}+O(J^{5})],
\end{equation}
where $A_{C} := -4\pi\left. \partial_{J} V(\chi_{c}(J)) \right|_{J=0}$ and $B_{C}:= -2\pi\left. \partial_{J}^{3} V(\chi_{c}(J)) \right|_{J=0}/3A_{C}$ which depend on the contour $C$ of the parameter control. Now $C$ is determined by the set of parameters $\alpha=(\hat{T}_{0},\hat{T}_{A})$ or $(\hat{\omega}_{C}, \hat{\omega}_{A}, \gamma_{R},\gamma_{A})$. 
With the aid of a numerical calculation for both cases, we obtain Figs. \ref{FT-beta} and \ref{FT-gw}, which show that $B_{C}$ is not negligibly small. 
This implies the geometrical current or the BSN curvature generates non-Gaussian fluctuations. Our result is consistent with the previous results of Ref. \cite{watanabe2}.

\subsection{Relations among cumulants}
Let us discuss the relations among cumulants.
The cyclic fluctuation relation (\ref{FR-SB}) can be rewritten as the integral form
\begin{equation}\label{int-FR}
\left\langle e^{-A_{C}(J+B_{C}J^{3})} \right\rangle \simeq 1.
\end{equation}
We expand $n$-th cumulant with $A_{C}$ in Eq. (\ref{FR-SB}) as
\begin{equation} \label{J-A}
\langle J^{n} \rangle_{c} =\sum_{m}L_{nm}A_{C}^{m}/m!. 
\end{equation}
From Eqs. (\ref{int-FR}) and (\ref{J-A}), we obtain the violation of the  FDR as
\begin{equation}\label{FDR}
2L_{11}-L_{20}+2B_{C}(L_{31}+3L_{11}L_{20}-L_{40})=0
\end{equation}
as the balance of terms of order $O(A_{C}^{2})$.
Similaly, we also obtain the violation of the nonlinear relation
\begin{equation}\label{nonlinear}
L_{12}-L_{21}+B_{C}(L_{32}+3L_{12}L_{20}+6L_{11}L_{21}-2L_{41})=0
\end{equation}
as the balance of terms of order $O(A_{C}^{3})$.
Note that we can obtain more relations as the balance of terms of any order $O(A_{C}^{n})$. The derivations of Eqs.(\ref{FDR}) and (\ref{nonlinear}) are given in Appendix \ref{sec:derivation-FDR}.
The violations of the conventional relations in Eqs. (\ref{FDR}) and (\ref{nonlinear}) are caused by the non-Gaussianity $B_{C}$ which originates from the BSN curvature.

\subsection{Instantaneous fluctuation relation}
\begin{figure}[htb]
\begin{center} %センタリングする
    \includegraphics[width=8cm]{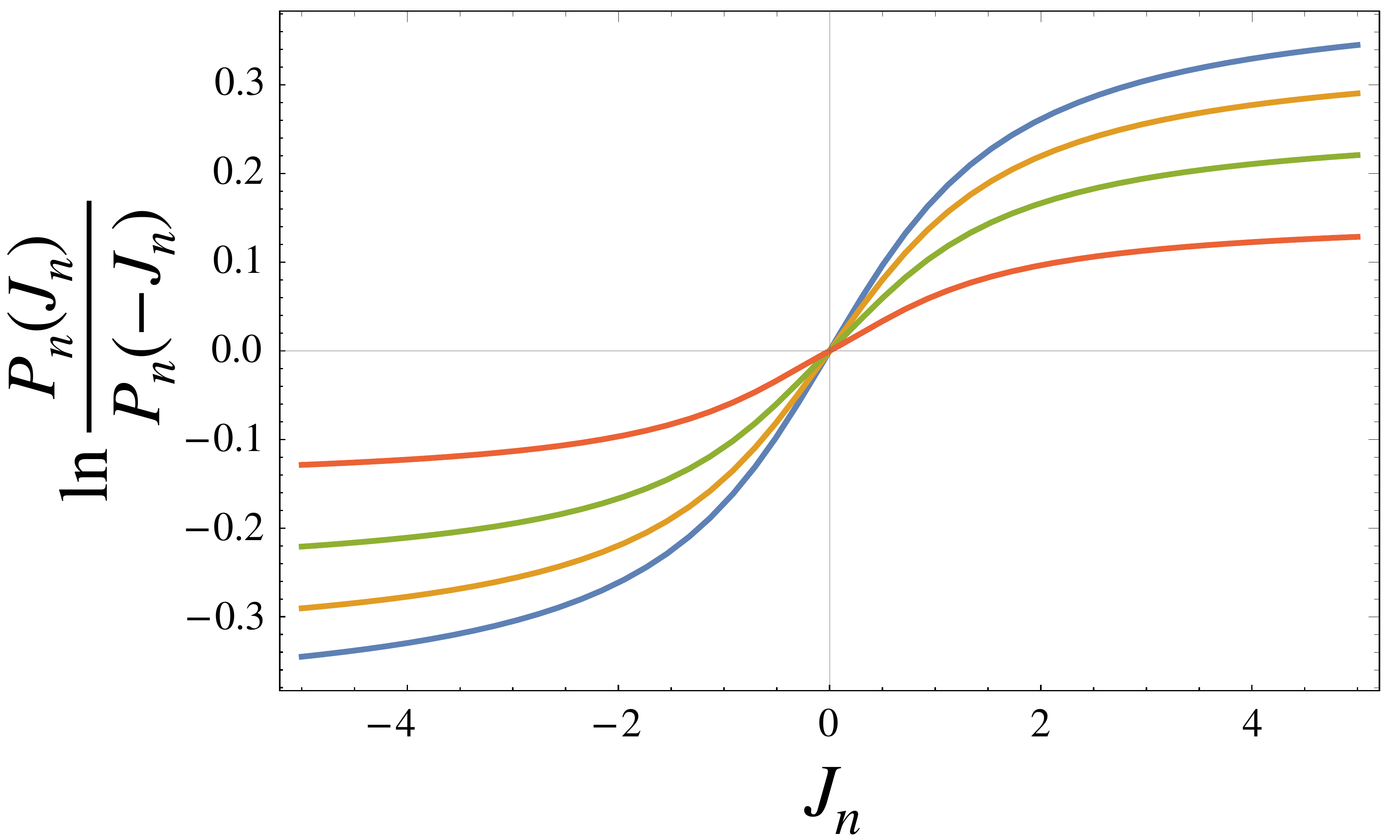}
    \caption{Plot of the instantaneous fluctuation relation at $\theta_{n}=\pi$ with $A_{n}=0$. $\hat{T}_{0}=0.5$ and  $\hat{T}_{A}=0.1,0.2,0.3,0.4$ (blue, yellow, green and red, respectively).} %タイトルをつける
    \label{ins-FT-SB} %ラベルをつけ図の参照を可能にする
  \end{center}
\end{figure}

Let us discuss the instantaneous fluctuation relation in this subsection. 
Here, we control temperatures in the right and the left reservoirs as in Eqs. (\ref{betaL}) and (\ref{betaR}). 
At $\theta_{n}=\pi n$ and $\mathcal{A}_{n}=0$ the geometrical contribution (the second term on the right hand side of Eq. (\ref{ins-FR})) is dominant. Its explicit behaviour is plotted as in Fig.\ref{ins-FT-SB}. 
This result implies that the fluctuation is highly non-Gaussian, in contrast to the conventional fluctuation theorem.

\section{Conclusion} \label{sec:conclusion}

In this paper, we derived the cyclic and the instantaneous fluctuation relations given in Eqs. (\ref{FR1}) and (\ref{ins-FR}), respectively, for adiabatic pumping processes.
We applied these results to the spin-boson model and clarified the existence of non-Gausianity as the geometric phase contribution in the fluctuation relations as in Eq. (\ref{FR-SB}) (Fig. \ref{FT-beta}). 
We confirmed that the non-Gaussianity $\hat{V}_{3}(\alpha)$ in Eq.(\ref{FR-SB}) is not small.
From the cyclic fluctuation relation, we obtained the relations among cumulants (\ref{FDR}) and (\ref{nonlinear}), which show the violation of the FDR and other conventional relations among cumulants. 
Our results indicate that the conventional fluctuation theorem should be extended to include non-Gaussian fluctuations if the geometric phase effect exists under cyclic modulation of parameters. 

Our future tasks are as follows: 
(i) Because our analysis is restricted to the adiabatic case, we will have to try to extend our analysis to the non-adiabatic case. 
If we restrict our interest to a two-level system like the spin-boson system, we can use the analytic solution of the generalized master equation (\ref{gcme}) \cite{fhht}.
(ii) Shortcuts to adiabaticity (STA) can be used for the non-adiabatic pumping in which a finite pumping current can be realized under finite speed modulation \cite{fhht}. 
Therefore, we expect that the universal work-fluctuation relation discussed by Funo et al. \cite{funo} can be generalized to include non-Gaussian fluctuations as mentioned in this paper. 
(iii) We can analyze the entropy production by a parallel method reported in Refs. \cite{sagawa, yuge2, nakajima2} in which the excess entropy production can be expressed by the geometric phase. 
(iv) We will have to discuss the linear response around a cyclic adiabatic state obtained in this paper by changing the modulation perturbatively. This linear response theory is expected to be different from that obtained from the Green-Kubo formula \cite{potanina}. 
(v) We have shown that Eqs. (\ref{sym-I}) and (\ref{fermion-affinity}) for dynamical part are held at least for the two-level spin-boson model, but we still do not have rigorous proof for the symmetry relation for general cases. 
Therefore we need further investigation on this problem. 
(vi) In this paper we have analyzed a classical system. 
To know quantum coherence effects, we need to analyze the Lindblad equation which has off-diagonal elements, which might induce nontrivial effect \cite{coherence}.

\section*{Acknowledgements} \label{sec:acknowledgements}
The authors thank Nigel Goldenfeld, Kazutaka Takahashi, Kiesuke Fujii, Ken Funo, Hiroyasu Tajima and Keiji Saito for fruitful discussions.
We also thank Ville Paasonen for his critical reading of this manuscript. 
This work is partially supported by a Grant-in-Aid of MEXT for Scientific Research (Grant No. 16H04025). 

\appendix

\section{Adiabatic approximation}\label{sec:ad-app}
Let us assume that the solution $|p(\theta,\chi)\rangle$ of the generalized master equation (\ref{gcme}) is parallel to the right eigenvector $|r(\theta, \chi)\rangle$ as
\begin{equation}
|p(\theta,\chi)\rangle = C(\theta,\chi)|r(\theta, \chi)\rangle,
\end{equation}
where the function $C(\theta,\chi)$ will be determined later. By using the generalized master equation (\ref{gcme}) and the normalization condition $\langle l(\theta, \chi)|r(\theta, \chi)\rangle =1$, we get
\begin{equation}\label{eq-of-c}
\partial_{\theta}C(\theta,\chi) =
C(\theta,\chi) \left[\epsilon^{-1}\lambda(\theta,\chi) - v(\theta,\chi)\right],
\end{equation}
where $v(\theta,\chi)=\langle l(\theta, \chi)|\partial_{\theta}|r(\theta, \chi)\rangle$.
Equation (\ref{eq-of-c}) can be solved as
\begin{align}
C(\theta,\chi) &= 
C(0,\chi) \exp 
\left[\int_{0}^{\theta} d\theta' [\epsilon^{-1}\lambda(\theta', \chi) -v(\theta',\chi)\right].
\end{align}
From the normalization condition $\langle 1|p(0,\chi)\rangle =\langle 1|p(0)\rangle=1$, we get
$C(0,\chi) = \langle 1| r(0,\chi)\rangle^{-1}$.
Therefore, we obtain Eqs.(\ref{ad}).

\section{Pumping current}\label{sec:current}
Although we are not interested in the average current for the adiabatic pumping process, it is useful to write its explicit form for the convenience to compare our results with the results in the literature. The average current can be decomposed to two parts as $\langle J \rangle 
= \langle J \rangle^{\mathrm{dyn}} + \langle J \rangle^{\mathrm{geo}}$.
The first part is the dynamic current expressed as
\begin{equation}
    \langle J \rangle^{\mathrm{dyn}}
    = \frac{1}{2\pi} \int_{0}^{2\pi} d\theta J^{\mathrm{st}}(\theta),
\end{equation}
where $J^{\mathrm{st}}(\theta):=\left.\partial_{i\chi}\lambda(\theta,\chi)\right|_{\chi=0}$ is the instantaneous steady current.
The second part is the geometrical current expressed as
\begin{equation}\label{BSNcurvature}
    \langle J \rangle^{\mathrm{geo}}
    =-\frac{\epsilon}{2\pi}\iint_{S} d\alpha_{m} d\alpha_{n} F_{mn}(\bm{\alpha}),
\end{equation}
where $F_{mn}(\bm{\alpha}):= \left.\partial_{i\chi}\mathcal{F}(\bm{\alpha},\chi)\right|_{\chi=0}$ is the BSN curvature \cite{sinitsyn1,sinitsyn2} in the parameter space.
Note that even if the average bias is zero such that $\langle J \rangle^{\mathrm{dyn}} = 0$, $\langle J \rangle^{\mathrm{geo}}$ is generally not zero.

\section{Derivation of Eq.(\ref{pdf})}\label{sec:calculation}
The current distribution is given in Eq. (\ref{pdf}).
The LDF is given in (\ref{ldf}).
We expand $\Lambda(\chi)$ with respect to $i\chi$ around $\chi=\chi_{c}(J)$ as
\begin{align}
\Lambda(\chi) &=\Lambda(\chi_{c}) +\Lambda^{(1)}(\chi_{c})(i\chi-i\chi_{c}) 
\notag \\
&+\frac{1}{2}\Lambda^{(2)}(\chi_{c})(i\chi-i\chi_{c})^{2} +O((i\chi-i\chi_{c})^{3}),
\end{align}
where we have introduced $\Lambda^{(n)}(\chi_{c}(J)):= \left. \partial_{i\chi}^{n}\Lambda(\chi)\right|_{\chi=\chi_{c}(J)}$. 
Then, we get
\begin{equation}
i\chi J -\Lambda(\chi)\simeq I(J) + \frac{1}{2}\Lambda^{(2)}(\chi_{c}(J))(\chi-\chi_{c}(J))^{2}.
\end{equation}
Therefore we obtain Eq.(\ref{pdf}) as
\begin{align}
P_{\epsilon}(J) 
&\simeq e^{-\frac{2\pi}{\epsilon}I(J)} \notag \\
&\times\int_{-\infty}^{\infty}\frac{d\chi}{\epsilon}
e^{-\frac{\pi}{\epsilon}\Lambda^{(2)}(\chi_{c}(J))(\chi-\chi_{c}(J))^{2}} e^{-2\pi V(\chi)} \notag \\
&\simeq \frac{1}{\sqrt{\epsilon \Lambda^{(2)}(\chi_{c}(J))}}
e^{-\frac{2\pi}{\epsilon}[I(J)+\epsilon V(\chi_{c}(J))]}.
\end{align}

\section{Derivation of Eq.(\ref{decomposition})}
\label{sec:decomposition}
In this appendix, we explain the details of the derivation of Eqs. (\ref{decomposition}) and (\ref{ins-pdf}).
From the definition of $P_\epsilon(J)$ in Eq. (\ref{pdf}) we can write
\begin{align}\label{p1}
&P_{\epsilon}(J)
=\frac{1}{\epsilon} \int_{-\infty}^{\infty} d\chi e^{-\frac{2\pi}{\epsilon}[ i\chi J -G_{\epsilon}(\chi) ]} 
\notag \\
&= \frac{1}{\epsilon} \int_{-\infty}^{\infty} d\chi \exp \left[ -\frac{2\pi}{\epsilon}\left\{ i \chi J - \frac{1}{2\pi} \int_{0}^{2\pi} d\theta g(\chi, \theta) \right\} \right] 
\end{align}
where  $g(\theta,\chi) := \lambda(\theta,\chi) - \epsilon v(\theta,\chi)$. 
As mentioned in Sec. \ref{subsec:inst}, we discretize variables in the interval $[0,2\pi]$ into $N$ pieces. Then, we rewrite Eq.(\ref{p1}) as 
\begin{widetext}
\begin{align}\label{p2}
P_{\epsilon}(J)
&= \frac{1}{\epsilon} \lim_{N \to \infty} \int_{-\infty}^{\infty} d\chi \exp \left[ -\frac{2\pi}{\epsilon} \left\{ i\chi J - \frac{1}{N} \sum_{n=1}^{N} g_{n}(\chi) \right\} \right]
\notag \\
&= \frac{1}{\epsilon} \lim_{N \to \infty} \int_{-\infty}^{\infty} d\chi e^{-\frac{2\pi}{\epsilon} i\chi J} \prod_{n=1}^{N} e^{\frac{2\pi}{\epsilon N}g_{n}(\chi)} \notag \\
&= \frac{1}{\epsilon} \lim_{N \to \infty} \int_{-\infty}^{\infty} d\chi e^{- \frac{2\pi}{\epsilon}i\chi J} 
\prod_{n=1}^{N} 
\int_{-\infty}^{\infty} d\chi_{n} \delta(\chi-\chi_{n}) e^{\frac{2\pi}{\epsilon N}g_{n}(\chi_{n})},
\end{align}
where we have used $g_{n}(\chi) = \int_{-\infty}^{\infty} d\chi_{n} \delta(\chi-\chi_{n}) g_{n}(\chi_{n})$.
By using $\delta(x) = \frac{1}{2\pi} \int_{-\infty}^{\infty} dk \; e^{ikx}$,  Eq. (\ref{p2}) can be rewritten further as
\begin{align}
P_{\epsilon}(J)
&= \frac{1}{\epsilon} \lim_{N \to \infty} \int_{-\infty}^{\infty} d\chi e^{-\frac{2\pi}{\epsilon} i\chi J} 
\prod_{n=1}^{N} 
\int_{-\infty}^{\infty} d\chi_{n} 
\int_{-\infty}^{\infty} \frac{dJ_{n}}{\epsilon N} e^{\frac{2\pi}{\epsilon N} i (\chi - \chi_{n}) J_{n}}
e^{\frac{2\pi}{\epsilon N}g_{n}(\chi_{n})} 
\notag \\
&= \frac{1}{\epsilon} \lim_{N \to \infty}
\prod_{n=1}^{N}\int_{-\infty}^{\infty} \frac{dJ_{n}}{\epsilon N}
 \int_{-\infty}^{\infty} d\chi \exp\left[ - \frac{2\pi}{\epsilon N} i \chi \left(J- \frac{1}{N} \sum_{n=1}^{N} J_{n} \right) \right]
\prod_{n=1}^{N}\int_{-\infty}^{\infty} d\chi_{n} 
 e^{- \frac{2\pi}{\epsilon N}(i \chi_{n} J_{n} - g_{n}(\chi_{n}))}
\notag \\
&= \frac{1}{\epsilon} \lim_{N \to \infty} \prod_{n=1}^{N}\int_{-\infty}^{\infty}  dJ_{n} 
\int_{-\infty}^{\infty} d\chi \exp\left[ -\frac{2\pi}{\epsilon N} i\chi \left(J- \frac{1}{N} \sum_{n=1}^{N} J_{n} \right) \right]
\prod_{n=1}^{N}  
\int_{-\infty}^{\infty} \frac{d\chi_{n}}{\epsilon N} e^{- \frac{2\pi}{\epsilon N}(i \chi_{n} J_{n} - g_{n}(\chi_{n}))}
\notag \\
&= \lim_{N \to \infty} N \prod_{n=1}^{N}\int_{-\infty}^{\infty} dJ_{n} \delta \left(J - \frac{1}{N} \sum_{n=1}^{N} J_{n} \right)   \prod_{n=1}^{N} p_{n}(J_{n}) .
\end{align}
\end{widetext}
Here we obtain the explicit expression of $p_{n}(J_{n})$ as Eq.(\ref{ins-pdf}).

\section{Detail of the spin-boson model}\label{sec:spin-boson-detail}

The eigenvalue $\lambda(\theta,\chi)$ of $\hat{K}(\bm{\alpha}(\theta),\chi)$ and corresponding eigenvectors $\langle l(\theta,\chi)|$, $|r(\theta,\chi)\rangle$ are given as
\begin{align}
&\lambda(\theta,\chi) 
= 
-\frac{k_{01}(\theta) + k_{10}(\theta) }{2}  \\ \notag
&+ \sqrt{ \left( \frac{ k_{01}(\theta) - k_{10}(\theta)}{2} \right)^{2} + k_{01}(\theta,\chi) k_{10}(\theta,\chi) },
\\
&\langle l(\theta,\chi)| = \left(1,   \frac{\lambda(\theta,\chi) + k_{10}(\theta) }{ k_{10}(\theta,\chi) } \right),
\\
&|r(\theta,\chi) \rangle = \frac{1}{ c(\theta,\chi) } \left(
\begin{array}{c}
\displaystyle
1 
\\
\displaystyle
\frac{\lambda(\theta,\chi) + k_{10}(\theta) }{ k_{01}(\theta,\chi) }
\end{array}
\right),
\\
&c(\theta,\chi) := 1 +\frac{ (\lambda(\theta,\chi) + k_{10}(\theta) )^{2} }{ k_{01}(\theta,\chi) k_{10}(\theta,\chi) }.
\end{align}

\section{Numerical check of Eq.(\ref{sym-I}) for the spin-boson model}\label{spin-boson-sym}
We derive the LLGC symmetry for $\lambda(\theta,\chi)$ in the spin-boson model.
The eigenvalue $\lambda(\theta,\chi)$ depends on $\chi$ through $k_{01}(\theta,\chi) k_{10}(\theta,\chi)$.
For any $\chi$, we obtain
\begin{align}
    0=& k_{01}(\theta,\chi)k_{10}(\theta,\chi) \notag \\
    &-k_{01}(\theta,-\chi+i\mathcal{A}(\theta))k_{10}(\theta,-\chi+i\mathcal{A}(\theta)) \notag \\
    =&(k_{01}^{L}(\theta)k_{10}^{R}(\theta)-k_{01}^{R}(\theta)k_{10}^{L}(\theta)e^{\mathcal{A}(\theta)}) \notag \\
    &\times (e^{i\chi} + e^{-i\chi}e^{-\mathcal{A}(\theta)}),
\end{align}
where
\begin{align}\label{spin-boson-ins-affinity}
    \mathcal{A}(\theta)
    &=\ln\frac{k_{01}^{R}(\theta)k_{10}^{L}(\theta)}{k_{01}^{L}(\theta)k_{10}^{R}(\theta)} =\ln\frac{n_{R}(\theta)(1+n_{L}(\theta))}{n_{L}(\theta)(1+n_{R}(\theta))}.
\end{align}
This reduces to $\mathcal{A}(\theta) =\hbar \omega_{0}(\beta_{R}(\theta)-\beta_{L}(\theta))$.
Therefore we obtain the LLGC symmetry $\lambda(\theta,\chi)=\lambda(\theta,-\chi+i\mathcal{A}(\theta))$.
This leads the symmetry relation (\ref{ins-sym}) for the instantaneous LDF $I_{n}(J_{n})$.

\begin{figure}[htbp]
    \centering
    \includegraphics[width=8cm]{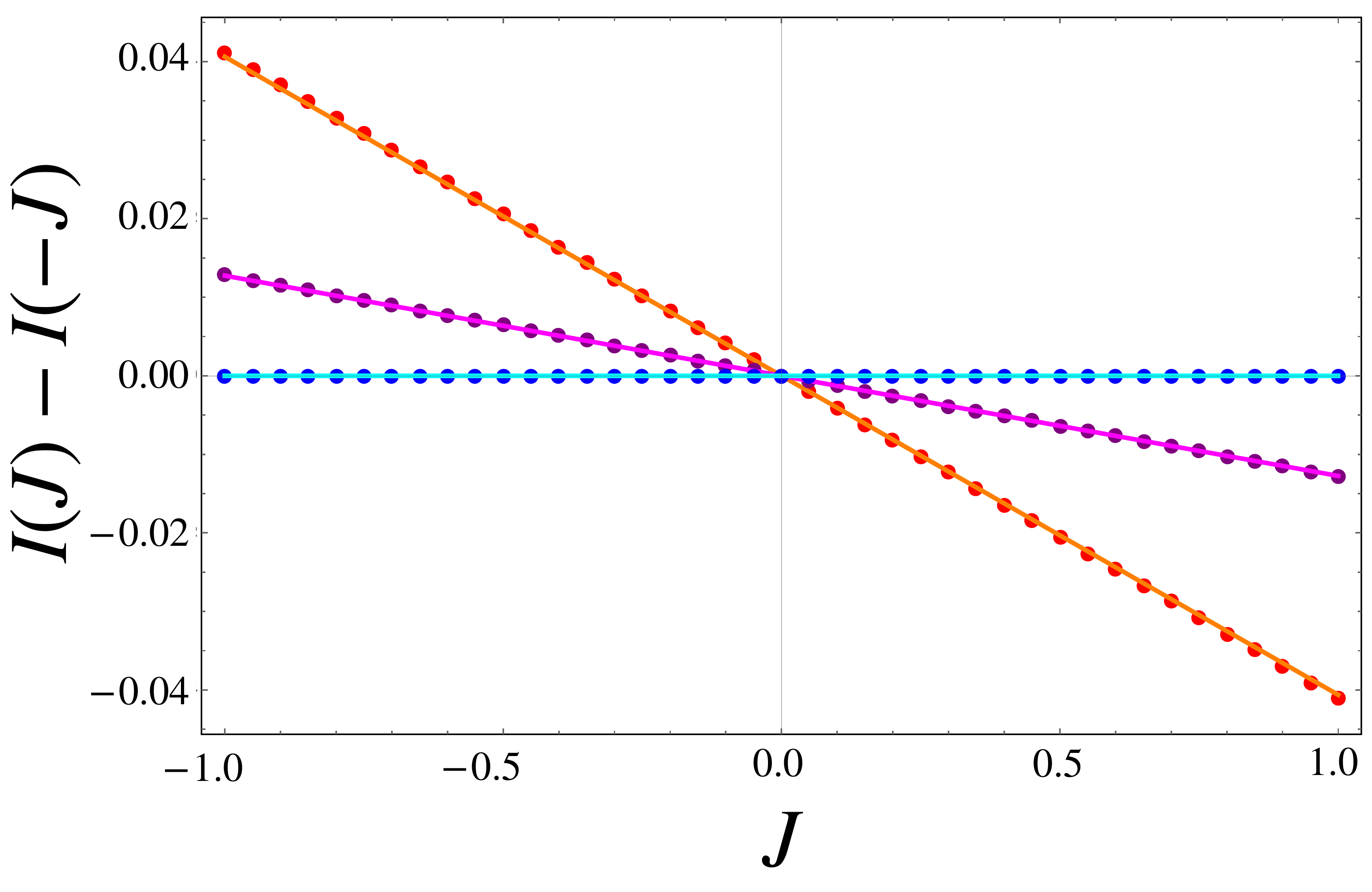}
    \caption{The verification of the symmetry relation of the large deviation function, Eq. (\ref{ldf}), in the spin-boson model. The points are the plots of $I(J)-I(-J)$ at $(\hat{T}_{R0},\hat{T}_{RA})=(5,4)$ and $(\hat{T}_{L0},\hat{T}_{LA})=(10,9), (6,5), (5,4)$ (their colors are red, purple, blue, respectively). The solid lines are the plots of $-\mathcal{A}J$ at $(\hat{T}_{R0},\hat{T}_{RA})=(5,4)$ and $(\hat{T}_{L0},\hat{T}_{LA})=(10,9), (6,5), (5,4)$ (their colors are orange, magenta and cyan, respectively). }
    \label{sym_numerical}
\end{figure}

When we consider the cyclic modulation in the limit $\epsilon\to 0$, 
the rate function $I(J)$ can be evaluated only from the dynamical part. 
We expect that the logarithmic form of Eq. (\ref{spin-boson-ins-affinity}) satisfies Eq. (\ref{fermion-affinity}) if we replace the numerator and the denominator in the right hand side of Eq. (\ref{spin-boson-ins-affinity}) by its cyclic average.
To confirm its validity, we numerically check the validity of Eq. (\ref{sym-I}).
We control the temperature of the left and right reservoirs as
\begin{align}
    &\hat{T}_{L}(\theta) = \hat{T}_{L0} + \hat{T}_{LA}\cos(\theta+\pi/4), \\
    &\hat{T}_{R}(\theta) = \hat{T}_{R0} + \hat{T}_{RA}\sin(\theta+\pi/4).
\end{align}
From the numerical calculation of the left and right hand sides of Eq. (\ref{sym-I}) in the spin-boson model, we have confirmed the symmetry relation (\ref{sym-I}) numerically (Fig. \ref{sym_numerical}).
Note that, from our numerical calculation, $\Lambda(\chi)=\Lambda(-\chi+i \mathcal{A})$ seems to hold. 

\section{Derivation of Eqs. (\ref{FDR}) and (\ref{nonlinear})}\label{sec:derivation-FDR}
Eq. (\ref{int-FR}) can be rewritten as 
\begin{equation}
    0= \sum_{n=1}^{\infty}(-A_{C})^{n}\langle (J+B_{C}J^{3})^{n}\rangle.
\end{equation}
Each moment $\langle J^{n} \rangle$ can be rewritten by cumulants $\langle J^{n}\rangle_{c}$ as
\begin{align}
    &\langle J\rangle = \langle J\rangle_{c}, \\
    &\langle J^{2}\rangle = \langle J^{2}\rangle_{c}+\langle J\rangle_{c}^{2}, \\
    &\langle J^{3}\rangle = \langle J^{3}\rangle_{c}+3\langle J^{2}\rangle_{c}\langle J\rangle_{c}+\langle J\rangle_{c}^{3}.
\end{align}
By using Eq. (\ref{J-A}), we obtain
\begin{align}
    &0=-A_{C}L_{10} \notag \\
    &+A_{C}^{2}\left[ -L_{11}+\frac{1}{2}L_{20}-B_{C}\left(L_{13}+3L_{20}L_{11}-L_{40}\right)\right] \notag \\
    &+A_{C}^{3}\left[ -\frac{L_{12}}{2}+\frac{L_{21}}{2} -B_{C}
    \left(\frac{L_{32}}{2}+3L_{21}L_{11}+\frac{3}{2}L_{20}L_{12}-L_{41}
    \right) \right] \notag \\
    &+O(A_{C}^{4}).
\end{align}
Then, we obtain Eqs.(\ref{FDR}) and (\ref{nonlinear}) as the balance of terms of order $O(A_{C}^{2})$ and $O(A_{C}^{3})$, respectively.

\end{document}